\documentclass[aps,letterpaper,amsmath,amssymb,reprint,superscriptaddress]{revtex4-1}
\usepackage{graphicx}
\usepackage{microtype}
\usepackage[utf8]{inputenc} 
\usepackage{float}
\usepackage{bm}

\usepackage{color}

\begin{document}
\title{Ultra-low switching current density in all-amorphous W-Hf / CoFeB / TaOx films}
\author{K. Fritz}
\email{kfritz@physik.uni-bielefeld.de}
\affiliation{Center for Spinelectronic Materials and Devices, Department of Physics, Bielefeld University, D-33501 Bielefeld, Germany} 

\author{L. Neumann}
\affiliation{Argelander Institute for Astronomy, Department of Physics, Bonn University, D-53121 Bonn, Germany}

\author{M. Meinert}
\email{markus.meinert@tu-darmstadt.de}
\affiliation{Department of Electrical Engineering and Information Technology, Technical University of Darmstadt, Merckstraße 25, D-64283 Darmstadt, Germany}

\date{\today}

\begin{abstract}
We study current-induced deterministic magnetization switching and domain wall motion via polar Kerr microscopy in all-amorphous W$_{66}$Hf$_{34}$/CoFeB/TaO$_\text{x}$ with perpendicular magnetic anisotropy and large spin Hall angle. Investigations of magnetization switching as a function of in-plane assist field and current pulse-width yield switching current densities as low as $3\times 10^{9}$ A/m$^2$. We accredit this low switching current density to a low depinning current density, which was obtained from measurements of domain wall displacements upon current injection. This correlation is verified by investigations of a Ta/CoFeB/MgO/Ta reference sample, which showed critical current densities of at least one order of magnitude larger, respectively. 
\end{abstract}

\maketitle

\section{Introduction}

Switching of ferromagnetic thin films by means of the spin Hall effect (SHE) in heavy metal/ferromagnetic (HM/FM) bilayers with perpendicular magnetic anisotropy (PMA) has been studied intensively over the last years \cite{Miron2011,Avci2012,Thiaville2012,Liu2012,Pai2012,Haazen2013,Martinez2013,Zhang2014,Cubukcu2014,Garello2014,LoConte2014,Yu2014,Hao2015,Prenat2016,Neumann2016}. In these systems, the switching current density $j_\text{sw}$ is typically in the order of $10^{10}$ to 10$^{11}$\,A/m$^2$. Recent work suggests that the SHE switching is limited by the depinning of domain walls \cite{Lee2014}. Therefore, a decrease of the depinning current density $j_\text{dep}$ should result in a reduced $j_\text{sw}$. Micromagnetic simulations \cite{Kim2017} show that the pinning of domain walls depends on the ratio of the domain wall width $\pi\Delta_\mathrm{DW}$ to the average grain size $\langle L \rangle$, where the maximum pinning strength is found for $\pi \Delta_\mathrm{DW} \approx \langle L \rangle$. For sufficiently small grains ($ \langle L \rangle  \ll \pi \Delta_\mathrm{DW}$), local variations of the anisotropy, which give rise to pinning, are averaged out over the domain wall width, resulting in weak pinning. Pinning is thus minimized in single crystal films with $\langle L \rangle \gg \pi \Delta_\mathrm{DW}$ or in nanocrystalline/amorphous films with $\langle L \rangle \ll \pi \Delta_\mathrm{DW}$. This hypothesis is supported by a report on reduced pinning in nanocrystalline W/CoFeB/MgO thin films, prepared via sputter deposition at high deposition rates \cite{Jaiswal2017}, and a similar observation of weak pinning of skyrmions \cite{Legrand2017}.
Based on this hypothesis, in the present work we investigate the switching current densities in an all-amorphous system and study its relation to domain wall pinning. For this purpose, we continue our investigations of a W$_{x}$Hf$_{x-1}$ 8 nm / CoFeB 3 nm / TaO$_\text{x}$ 2 nm system \cite{Fritz2018}, which exhibits a phase transition from a segregated phase mixture to an amorphous alloy for $x \leq 0.7$. Due to the accompanying jump in resistivity, the SHA shows a pronounced maximum of $\theta_\text{SH} = -0.2$ at the phase transition. For the amorphous compositions, XRD patterns indicate at most some local order with a coherent scattering length of $D_\text{z} \approx \left(0.9 \pm 0.2\right)$\,nm. In this system, PMA can be obtained by decreasing the FM layer thickness and post-annealing the sample. First observations of domain nucleation and expansion upon out-of-plane field application showed the formation of large domains with only few pinning sites. We understand this as a hint for weak pinning in the amorphous W-Hf, making this system interesting for investigation of a correlation between the lack of crystallinity and weak pinning. Finally, the large SHA of the amorphous W-Hf should allow for efficient current-induced magnetization switching. 

Here, we report on current-induced magnetization switching experiments (CIMS) and current-induced domain wall motion (CIDWM) experiments, performed on all-amorphous W$_{66}$Hf$_{34}$/CoFeB/TaO$_\text{x}$ with perpendicular magnetic anisotropy and large spin Hall angle via polar Kerr-microscopy. For comparison, we additionally investigate a reference system consisting of Ta/CoFeB/MgO/Ta which is known to exhibit stronger domain wall pinning. After preparation via sputter deposition, both systems were electrically and magnetically characterized to ensure comparability, and patterned for the observation of domain wall motion and magnetization switching. In case of the W-Hf/CoFeB/TaO$_\text{x}$, the maintenance of the amorphous phase during the sample processing was verified by comparing the sheet resistance before and after post annealing. In order to study the correlation between the switching current density and the depinning current density we conduct the same CIMS and CIMDW experiments on both systems. We obtain critical current densities of at least one order of magnitude lower for the W-Hf system, respectively. To support our conclusion that a reduced depinning current density will result in a reduced switching current density, we evaluate the influence of Joule heating and the energy barriers for nucleation and depinning of domains in both systems. Additionally, we perform measurements of the spin Hall effective fields in various configurations via harmonic response analysis to ensure that the difference in the critical current densities is not a result of significantly different spin-orbit torque efficiencies.

\section{Methods}

\subsection{Sample Preparation}

The thin films were grown in UHV magnetron (co-) sputtering systems at room temperature on thermally oxidized Si wafers and post annealed in a vacuum furnace with a pressure below $5\times 10^{-7}$\,mbar. W-Hf thin films were prepared via co-sputtering. Thermal stability of the amorphous phase was verified by comparing the sheet resistance, measured via four-probe technique, before and after annealing. A nominal tungsten content of 66\,\% was chosen for further investigations. All layers in this sample were deposited with an Ar working pressure of $2\times 10^{-3}$\,mbar. Perpendicular magnetic anisotropy (PMA) was obtained with a thin CoFeB layer and by post annealing the sample at 180\,$^\circ$C for 20\,min. The full stack is Si (001) / SiO$_\text{x}$ 50\,nm / W$_{66}$Hf$_{34}$ 8\,nm / Co$_{40}$Fe$_{40}$B$_{20}$ 0.85\,nm / TaO$_\text{x}$ 2.55\,nm / SiN 1.5\,nm. For comparison, a reference sample consisting of Si (001) / SiO$_\text{x}$ 50\,nm / Ta 8\,nm / Co$_{40}$Fe$_{40}$B$_{20}$ 1.1\,nm / MgO 1.8\,nm / Ta 1.5\,nm was prepared. The Ar working pressure during the sputter deposition was $1.2\times 10^{-3}$ and, for the MgO layer, $2.2\times 10^{-2}$\,mbar. Here, MgO allows for PMA with a thicker CoFeB layer. Longitudinal and polar magneto-optical Kerr effect (MOKE) measurements at room temperature were performed to characterize the magnetization of the samples. In order to ensure comparability of the two systems, the anisotropy that determines the domain wall width is set to a similar value. Therefore, the reference sample was post annealed for 20\,min at 280\,$^\circ$C to match the anisotropy of the W-Hf system. Maintenance of the amorphous CoFeB during the annealing was, again, verified by measurements of the sheet resistance. Additionally, the crystallinity of the samples was investigated via X-ray diffraction (XRD) with Cu $K_\alpha$ radiation in a diffractometer with Bragg-Bretano geometry. The measurements were performed with a sample offset to attenuate substrate peaks and clearly assign any measured peaks to the thin films. The saturation magnetization $M_\text{s}$ was determined via vibrating sample magnetometry (VSM).

The samples were patterned via electron beam lithography and ion beam milling into line bars and Hall bars, and Ta/Au contact pads were deposited via sputter deposition subsequently. The line bars with geometries of $50\, \times 6$\,$\mu$m$^2$ and $100\, \times 2$\,$\mu$m$^2$, were used for observation of domain dynamics and switching upon current injection. The Hall bars, consisting of $40\, \times 3$\,$\mu$m$^2$ and $30\, \times 3$\,$\mu$m$^2$ bars with contact lines, were used to determine the SOT efficiencies~$\xi$. 

\subsection{Observation of Domain Wall Motion and Magnetization Switching}

CIMS was observed with a homebuilt Kerr microscope utilizing the polar MOKE \cite{Hubert2000,McCord2015}, using a commercial Carl Zeiss metallurgical microscope. While simultaneously applying current pulses and static in-plane magnetic fields ($B_\text{max}=400$\,mT) parallel to the line bars, the brightness of the Kerr image was measured as a function of the applied current density. From the resulting hysteresis loops (cf. Fig. \ref{j_sw}(a)), $j_\text{sw}$ was obtained in dependence on the pulse width $\tau$ and longitudinal field $B_\text{x}$. The final $j_\text{sw}$ was calculated as a weighted mean from at least three individual measurements. For more detailed information about the error analysis and statistics we would like to refer to our previous work\cite{Fritz2018}. 

CIDWM was observed in differential Kerr images.  The differential images are stabilized in software via frame-cropping and FFT-based image registration \cite{Guizar-Sicairos2008}. Noise reduction is achieved with a convolutional neural network (FFDNet) \cite{Zhang2018,Tassano2019} running in real-time on a graphics processing unit. The domain wall velocity $v_\text{DW}$ was obtained from the displacement of a domain wall divided by the number of pulses $N$ and $\tau$ as a function of applied current density. In multi-pulse experiments ($N > 1$), the pulse on/off ratio was chosen small enough to ensure full cool-down of the line between pulses.

\subsection{Harmonic Response Analysis}

The motion and expansion of domain walls is caused by dampinglike (DL) and fieldlike (FL) effective fields $B_\text{DL/FL}$, which originate from the SHE due to an in-plane current injection. A measure to compare these fields is the spin-orbit torque efficiency
\begin{equation}
\xi_\mathrm{DL/FL}= \frac{2e}{\hbar} \frac{M_\text{s} t_\text{FM} B_\text{DL/FL}}{j_0} c_\text{AR}.
\end{equation}
The DL and FL effective fields were obtained from harmonic Hall voltage measurements. All measurements were conducted by injecting an in-plane AC current $I(t) = I_0 \sin(\omega t)$ into the Hall bars and simultaneously recording the in-phase first harmonic and out-of-phase second harmonic Hall voltages with a Zurich Instruments MFLI multidemodulator lock-in amplifier. From the resulting effective fields and the current density amplitude $j_0$, the presented SOT efficiencies $\xi_\mathrm{DL/FL}$ were determined as a weighted mean from multiple measurements, taking into account the aspect ratio of the Hall bars with a correction factor $c_\text{AR} $\cite{Neumann2018}. Because of the similar resistivities of all conducting layers (about 180 to 200\,$\mu\Omega$cm), no correction within a parallel resistor model had to be applied. However, we take into account the different orientations of magnetic moments in both the domains and the domain walls. In the latter case the moments lie, even though the sample has PMA, in the film plane and the resulting effective fields are not necessarily the same as for the out-of-plane oriented magnetic moments. To separate and compare the effective fields, we conduct the measurements in two geometries of the external magnetic field, which is applied during the measurements.

For the determination of the effective fields with out-of-plane geometry, representing the magnetic moments in the domains, a measurement scheme as described by Hayashi et al. \cite{Hayashi2014} is useful. Following this scheme, the SOT fields 
\begin{equation}
B_{x/y} = - 2 \dfrac{\left( b_{x/y} \pm 2 \chi b_{y/x} \right)}{1 - 4 \chi^2}
\end{equation}
were determined. The sign $\pm$ denotes the orientation of $\vec{M}$ along $\pm z$ and $\chi = R_\text{P} / R_\text{A}$ the ratio of the planar and anomalous Hall resistances, which are obtained from in-plane field rotation and out-of-plane field sweeps, respectively.  $b_{x/y} \equiv \left( \frac{\partial V_{2\omega}}{\partial H} / \frac{\partial^2 V_\omega}{\partial H^2} \right)$ were extracted from in-plane field sweeps longitudinal ($x$) and transverse ($y$) to the Hall bars, where the $x$ direction corresponds to the DL and the $y$ direction to the FL effective field. The field sweeps were performed in a vector magnet consisting of two coil pairs oriented along the x- and y-direction and a maximum in-plane field of $0.3$\,T.

\begin{figure}[b]
	\includegraphics[width=8.6cm]{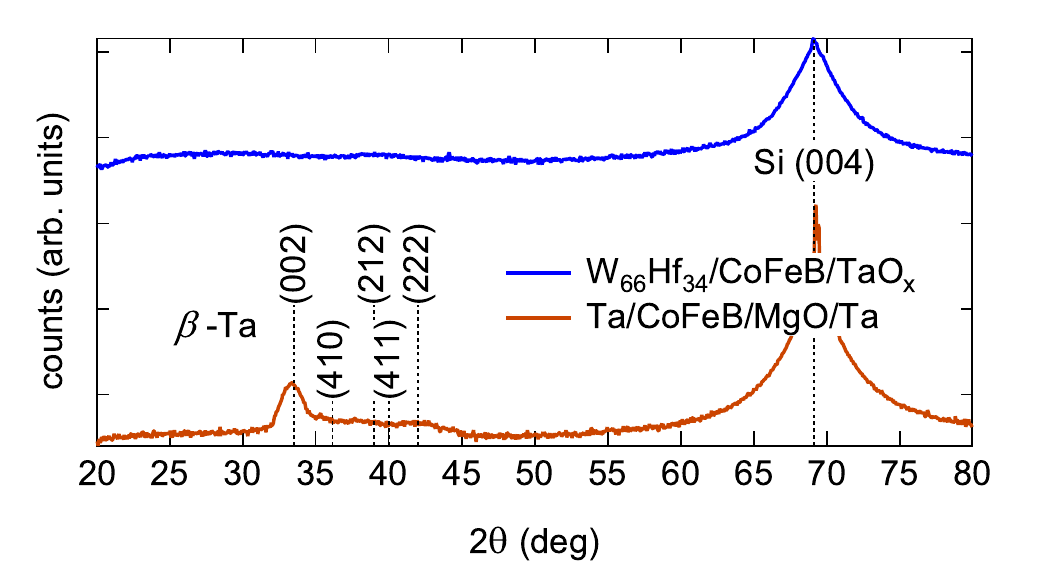}
	\caption{\label{XRD} XRD pattern of the annealed W$_{66}$Hf$_{34}$/CoFeB/TaO$_\text{x}$ and Ta/CoFeB/MgO/Ta sample stacks. The Si (004) Peak is attenuated by up to 400 times with respect to the specular geometry. The (forbidden) Si (002) peak is suppressed below the noise level of the measurement.}
\end{figure}

An in-plane measurement scheme identical to our previous work \cite{Fritz2018} was applied to obtain the effective fields that cause torques in the domain walls. Here, a dual Halbach cylinder array with a rotating magnetic field up to $B_\mathrm{ext} = 1.0$\,T (MultiMag, Magnetic Solutions Ltd.) was used for in-plane field rotation. The scheme is facilitated by the weak PMA in our samples. The current-induced effective SOT field amplitudes $B_\text{DL}$ and $B_\text{FL}$, associated with the DL and FL spin-orbit torques, were derived from the second harmonic Hall voltage rms value 
\begin{equation}
V_{2\omega} = \left( - \dfrac{B_\text{FL}}{B_\text{ext}} R_\text{P} \cos 2 \varphi - \dfrac{1}{2} \dfrac{B_\text{DL}}{B_\text{eff}} R_\text{A} + \alpha ' I_0 \right) I_\text{rms} \cos \varphi.
\end{equation}
The DL effective fields and the anomalous-Nernst contribution $\alpha ' I_0$ were separated by their dependence on the external field, see Ref. \cite{Fritz2018} for a detailed discussion of the method. Here, $\varphi$ is the in-plane field-angle with respect to the current direction and $B_\mathrm{eff} = -B_\mathrm{ani} + B_\mathrm{ext}$ is the effective magnetic field. We use the convention $B_\mathrm{ani} > 0$ for spontaneous PMA.

\section{Results and Discussion}

\begin{figure}[b]
	\includegraphics[width=8.4cm]{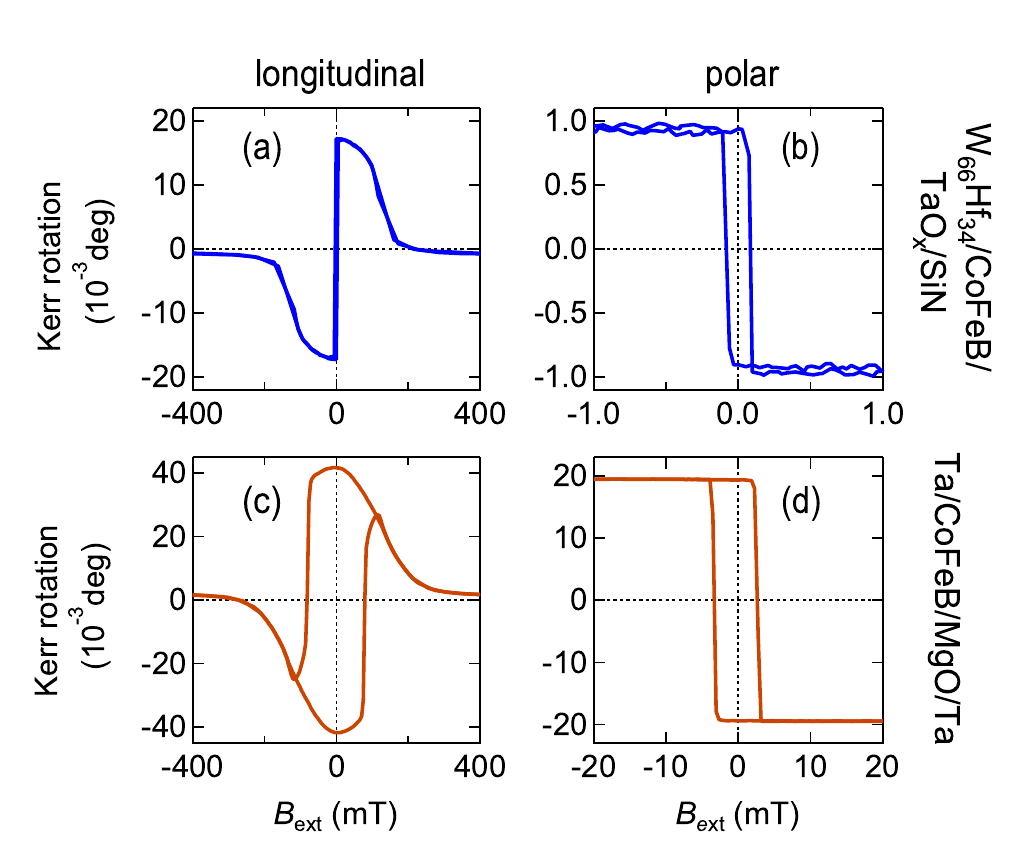}
	\caption{\label{l_p_MOKE} Longitudinal and polar MOKE measurements of the final (a) \& (b) W$_{66}$Hf$_{34}$/CoFeB/TaO$_\text{x}$/SiN and (c) \& (d) Ta/CoFeB/MgO/Ta sample stacks. }
\end{figure}

\begin{figure}[b]
	\includegraphics[width=8.6cm]{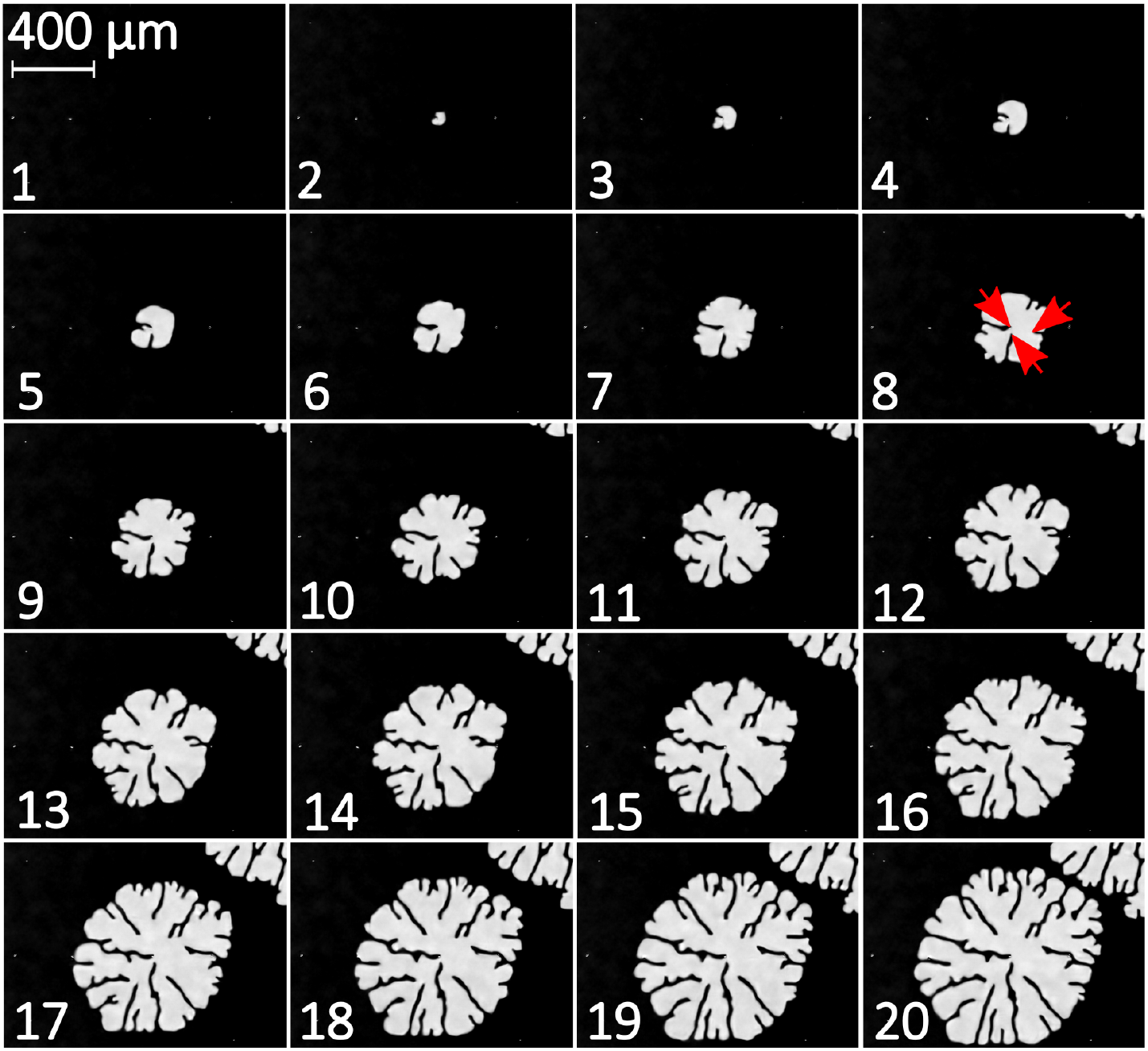}
	\caption{\label{switching_planefilm} Differential Kerr images of an opposing domain nucleated and expanded by out-of-plane field pulses in a W$_{66}$Hf$_{34}$/CoFeB/TaO$_\text{x}$/SiN film with $v_\text{DW} \approx 0.23$ mm/s. Each image was recorded after application of a single pulse with $B_\text{z}=100\, \mu$T and $\tau = 0.1$\,s. The red arrows in image 8 exemplary mark pinning sites.}
	\includegraphics[width=8.6cm]{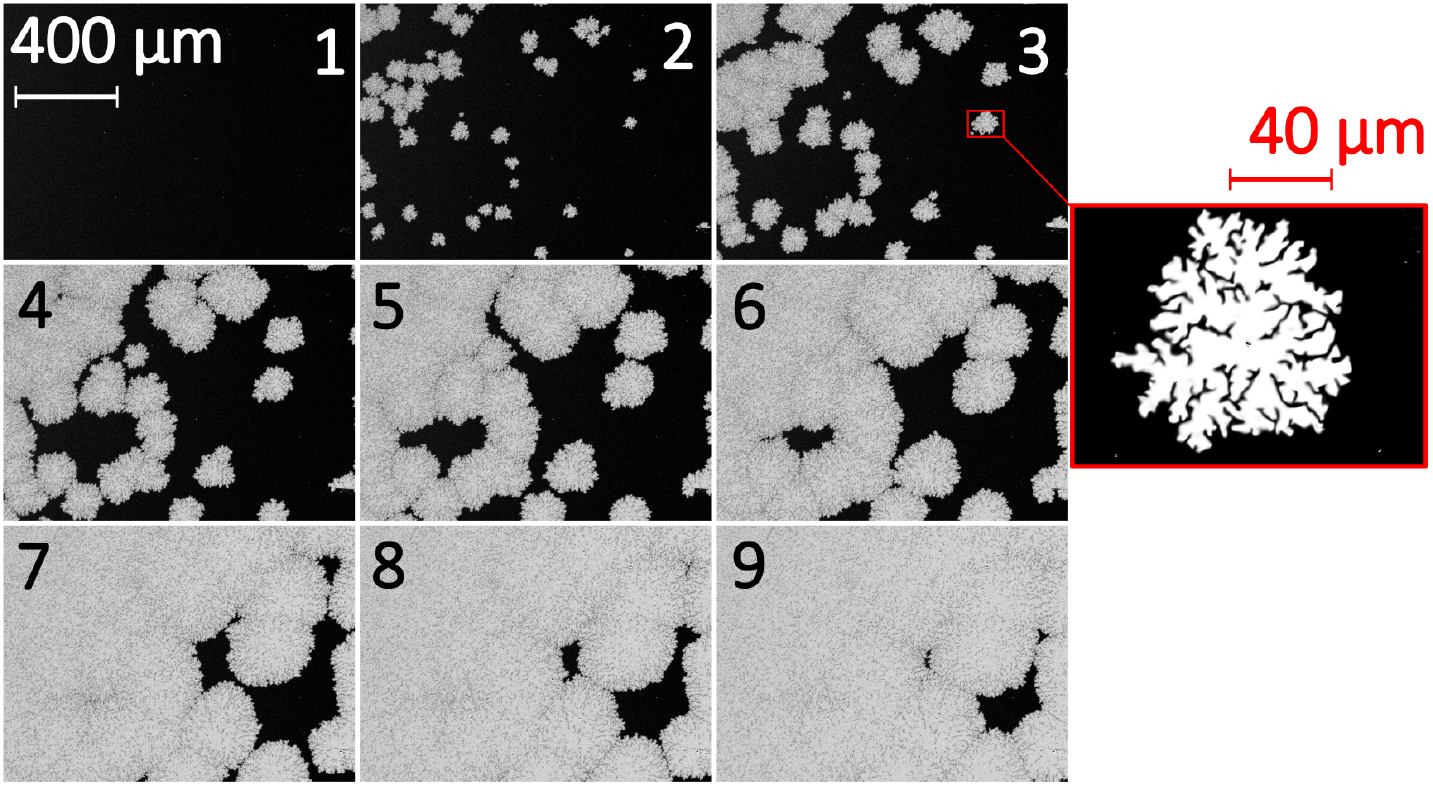}
	\caption{\label{planefilm_Ta}Differential Kerr images of opposing domains nucleated and expanded by out-of-plane field pulses in a Ta/CoFeB/MgO/Ta film with $v_\text{DW} \approx 0.23$ mm/s, each recorded after application of a single pulse with $B_\text{z}=2.7$\,mT and $\tau = 0.1$\,s. The red framed inset shows a domain nucleated at the marked area with the same field pulse parameters, observed with higher magnification.}
\end{figure}

In our previous work \cite{Fritz2018}, we reported on a large SHE in W-Hf/CoFeB/TaOx/SiN films which was observed as a result of the formation of an amorphous phase for a tungsten content below 70\%. Due to the accompanying jump in resistivity, the spin Hall angle shows a pronounced maximum of $\theta_\text{SH}=-0.2$ at the phase transition. This stoichiometry, however, showed a decrease in resistivity upon post annealing, indicating crystallization. In order to find a thermally stable amorphous composition with a SHA as large as possible, W-Hf thin films with nominal tungsten content of 62\% to 68\% were prepared. Thermal stability is verified in all samples and the measured high resistivities confirm equivalence to our previous work. To compensate for slight process instabilities during the deposition, a tungsten content of 66\% was chosen for further investigations. Due to the usage of TaO$_\text{x}$ instead of MgO, which is known to crystallize upon annealing, we assume to maintain the all-amorphous character of our sample stack throughout the sample processing. The final film stack has an effective resistivity of $\rho_\text{xx} = 185\,\mu \Omega$cm and a magnetization of $M_\text{s} = 684$\,kA/m. The low magnetization indicates the presence of a magnetic dead layer in the CoFeB, probably due to oxidation during the formation of the TaO$_\text{x}$ layer. The XRD pattern of the annealed sample stack, shown in Fig.\,\ref{XRD}, shows no evidence for a crystal structure or an atomic local order, indicating an all-amorphous system. From MOKE measurements, presented in Fig.\,\ref{l_p_MOKE}(a) and (b), an anisotropy field of $B_\text{ani}=123$\,mT and a very low coercive field of $B_\text{c} \approx 80\,\mu$T were determined. The latter allows for nucleation and expansion of domains in the  W$_{66}$Hf$_{34}$/CoFeB/TaO$_\text{x}$/SiN film with low out-of-plane field pulses, as shown in Fig.\,\ref{switching_planefilm} for field pulses with $B_\text{z}=100\,\mu$T and $\tau=0.1$\,s. The resulting domain only contains few macroscopically visible pinning sites, exemplarily highlighted with red arrows in Fig.\,\ref{switching_planefilm} image 8. Here, the initial magnetic orientation is pinned and causes the river-like structures in the radially expanding opposing white domain. From measurements of the growth of the domain after each field pulse, we estimate a domain wall velocity of $v_\text{DW} \approx 0.23$ mm/s, which corresponds to the creep regime as will be discussed further below. An analysis with the \textsc{ImageJ} plugin \textsc{Delaunay Voronoi} yields a pinning site mean distance of $\overline{d} = \left(144.5 \pm 68.5\right)\,\mu$m.

The Ta/CoFeB/MgO/Ta reference sample was post-annealed to match the anisotropy of the W$_{66}$Hf$_{34}$ sample, resulting in $B_\text{ani}=120$\,mT and $B_\text{c}=3$\,mT, determined from the MOKE measurements shown in Fig. \ref{l_p_MOKE} (c) and (d). The final sample stack has an effective resistivity of $\rho_\text{xx} \approx 180\,\mu \Omega$cm and a saturation magnetization of $M_\text{s} = 1088$\,kA/m. From the XRD pattern in Fig.\,\ref{XRD}, we infer the formation of $\beta$-Ta in this sample, which is in good agreement with its large resistivity \cite{Jiang2005}. The prominent peak at $2\theta = 33.35^\circ$ that corresponds to the $\beta$-Ta (002), allows for determination of a minimum length of coherent scattering $D_\text{z} \approx 6$\,nm via Scherrer's formula. The broad shoulder next to this peak could be caused by various other scattering planes of the $\beta$-Ta, some of which are marked in Fig.\,\ref{XRD}. In comparison to the W$_{66}$Hf$_{34}$ sample, here the application of out-of-plane field pulses results in the nucleation of many opposing domains with much narrower and denser river-like structures, as shown in Fig.\,\ref{planefilm_Ta}. In the interest of comparability, the domain growth was observed with the same magnification and domain wall velocity of $v_\text{DW} \approx 0.23$ mm/s, which was achieved with single pulses of $B_\text{z}=2.7$\,mT and $\tau=0.1$\,s. From the Delaunay triangulation we obtain a mean distance $\overline{d} = \left(9.6 \pm 5.0\right)\,\mu$m between the pinning sites.

\begin{figure}[b]
	\includegraphics[width=8.6cm]{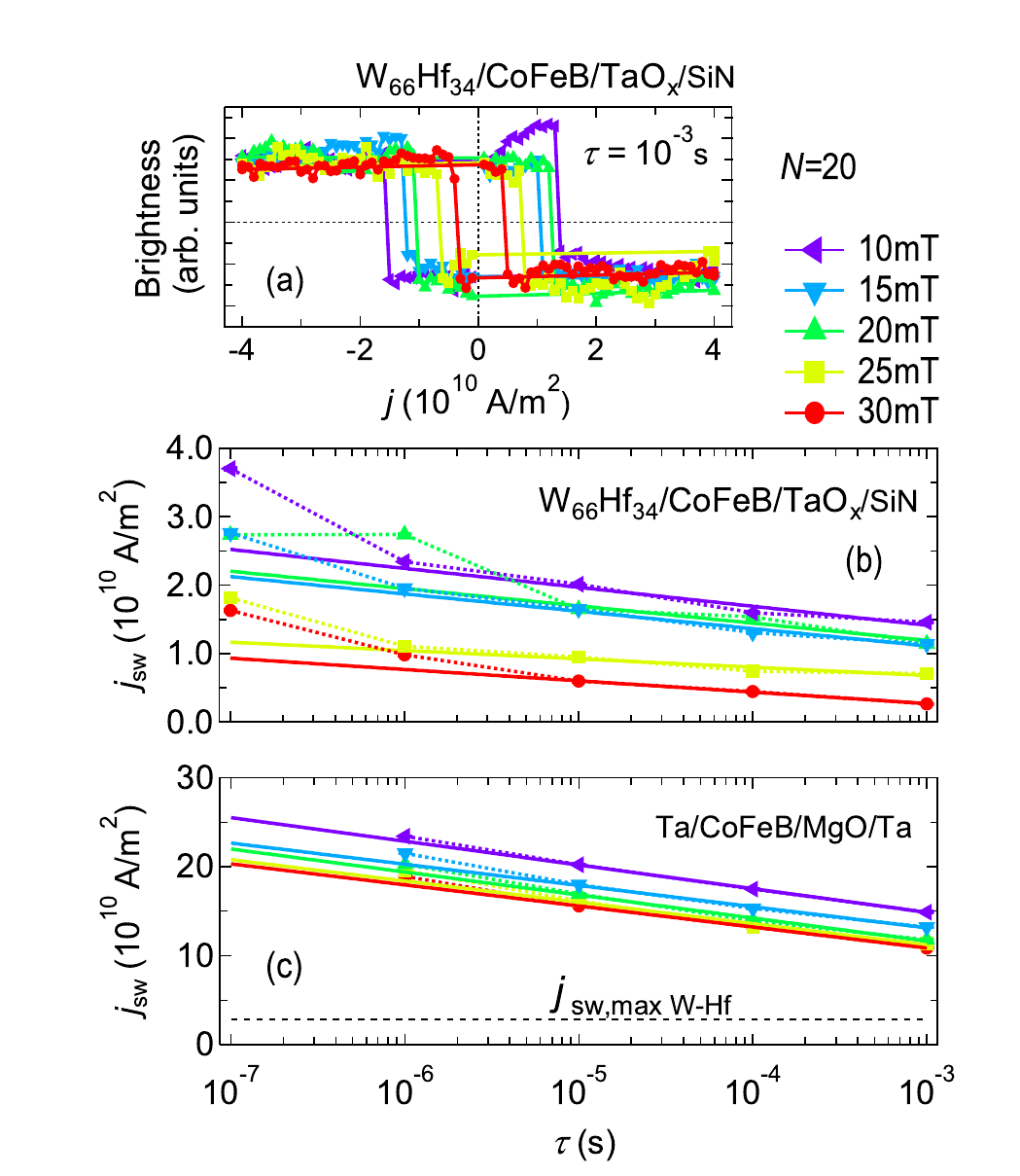}
	\caption{\label{j_sw} (a) Hysteresis loops recorded via Kerr microscopy showing the Kerr image brightness as a function of the applied current density for different in-plane fields $B_\text{x}$. Each data point was recorded after pulsing $N=20$ times with a pulse width of $\tau=1\times 10^{-3}$\,s. Switching current density in $50 \times 6$~$\mu$m$^2$ (b) W$_{66}$Hf$_{34}$/CoFeB/TaO$_\text{x}$/SiN and (c) Ta/CoFeB/MgO/Ta line bars as a function of pulse width for different in-plane fields obtained from the hysteresis loops. Data for $\tau = 10^{-7}$~s is missing for Ta, because an amplifier with impedance mismatch was used that distorted the short pulses. The solid lines show fits of Eq. \ref{NeelArrhnenius} to the experimental data.}
\end{figure}

In Fig.\,\ref{j_sw}(a) we show hysteresis loops obtained by observing CIMS in the $50\, \times 6$\,$\mu$m$^2$ line bars with varying in-plane longitudinal fields in the all-amorphous W$_\text{66}$Hf$_\text{34}$-based sample. Each data point was recorded after $N=20$ pulses with a pulse width $\tau=1\times 10^{-3}$\,s. Analogously, CIMS experiments were conducted with different pulse widths and the resulting j$_\text{sw}$ is displayed in Fig.\,\ref{j_sw}(b) as a function of $\tau$ for varying in-plane fields. We observe a decrease of $j_\text{sw}$ for increasing pulse width and longitudinal field, with ultra-low switching current densities in the range of $3\times 10^9 < j_\text{sw} < 2.8 \times 10^{10}$\,A/m$^2$. To the best of our knowledge, these are the lowest switching current densities reported so far in HM/FM bilayer systems and a similarly low switching current density was found only in an epitaxial system \cite{Li2019}. The same experiment was performed with the Ta reference sample. The resulting $j_\text{sw}$, presented in Fig.\,\ref{j_sw}(c), again shows a decrease with increasing field and pulse width and matches the results from other reports\cite{Lee2014,LoConte2014} found for similar layer stacks. However, with $1.2\times 10^{11} < j_\text{sw} < 3.0 \times 10^{11}$\,A/m$^2$ the switching current density of the Ta reference is up to 40 times larger than in the amorphous W$_{66}$Hf$_{34}$ sample.

\begin{figure}[b]
	\includegraphics[width=8.6cm]{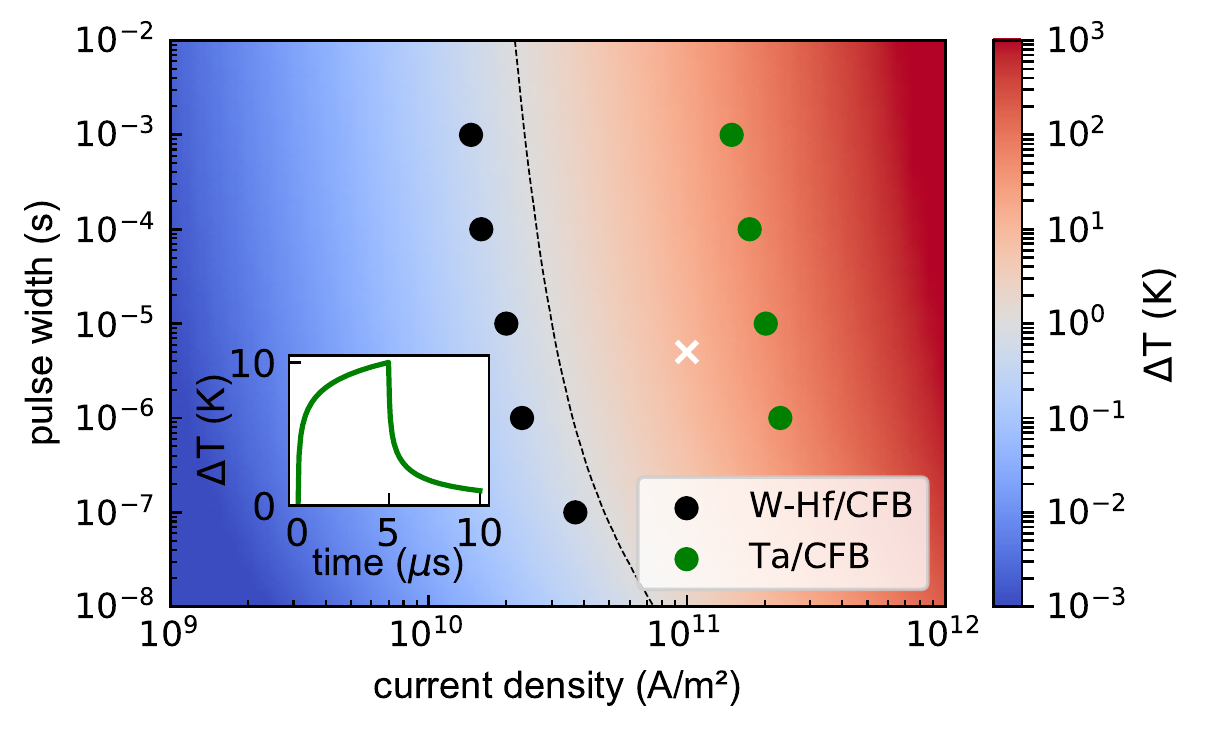}
	\caption{\label{heatmap_with_experimental_data_6um} Calculated temperature rise $\Delta T$ due to Joule heating as a function of pulse width $\tau$ and current density $j$ from a two-dimensional model (Eq. \ref{heatmapequation}) \cite{You2006}. The markers highlight the $\Delta T$ corresponding to the switching current densities obtained for the W$_{66}$Hf$_{34}$ and Ta sample with $B_\text{x}=10$~mT and the investigated pulse widths $\tau$, respectively. The dashed line marks $\Delta T = 1$~K. The inset shows the time-dependence of $\Delta T$ for $j=10^{11}$\,A/m$^2$ and $\tau = 5 \times 10^{-6}$\,s (white cross).}
\end{figure}

In order to rule out excessive Joule heating as a reason for the ultra-low switching current densities in the W$_{66}$Hf$_{34}$ sample, we calculate the temperature rise $\Delta T$ in the line bars. We use a two-dimensional model as derived by You \textit{et al.} \cite{You2006}, to obtain the maximum temperature rise at the end of a current pulse ($t=\tau$):
\begin{align}
\Delta T (t) &= c \frac{whj^2}{\pi \kappa_S \sigma} \Bigg[ \text{arcsinh}\left( \frac{2 \sqrt{\mu_S t\, }}{\alpha w} \right) \nonumber\\
&- \theta\left( t-\tau \right) \text{arcsinh}\left( \frac{2 \sqrt{\mu_S \left(t - \tau \right)}}{\alpha w} \right) \Bigg]
\label{heatmapequation}
\end{align}
Here, $\kappa_s$ and $\mu_s$ are the heat conductivity and the thermal diffusivity of the Si substrate and $c$ is a temperature scaling factor to take into account the $50$~nm SiO$_2$ layer on top of the substrate. The latter was determined from FEM simulations to be $c\approx1.45$. $\alpha = 0.5$ was chosen as proposed by You \textit{et al.}. As both samples are comparable in terms of film thickness and resistivity we estimate $\Delta T$ with $h=9$~nm, $\rho= 1/\sigma=180$~$\mu\Omega$cm, and $w=6$~$\mu$m as a function of $\tau$ and $j$, presented in Fig.\,\ref{heatmap_with_experimental_data_6um} with a logarithmic scale. The inset shows $\Delta T(t)$ for $j=10^{11}$\,A/m$^2$ and $\tau = 5 \times 10^{-6}$\,s. The black and green markers highlight the $\Delta T$ in the W$_{66}$Hf$_{34}$ and Ta line bars corresponding to the switching current densities found for $B_\text{x}=10$~mT and the investigated pulse widths $\tau$, respectively. The dashed line is the equi-temperature line with $\Delta T = 1$~K as a guide to the eye. We find a maximum temperature rise of $\Delta T_\text{Ta} \approx 50$~K in the Ta line bars, while in the W$_{66}$Hf$_{34}$ the temperature rise does not exceed $\Delta T_\text{W-Hf} \approx 0.6$~K. Therefore, the Joule heating is low in the W$_{66}$Hf$_{34}$ and can be neglected. In the Ta line bars, on the other hand, the Joule heating is more prominent and we rather underestimate the switching current densities here due to the additional thermal activation provided by the Joule heating. Next, we estimate the energy barrier $E_\text{B,0}$ for nucleation of an opposing domain. We start from the Néel-Arrhenius equation $\tau = \tau_0 \cdot \exp \left(E_\text{B} / k_\text{B}T\right)$ and assume a lowered energy barrier $E_\text{B} = E_\text{B,0} \cdot \left( 1- j_\text{c}/j_\text{c,0} \right)$ due to the effective fields that result from the current pulses. This leads to the following fit equation for the switching current density \cite{Koch2004}
\begin{equation}
j_\text{c} = j_\text{c,0} \left( 1 - \dfrac{k_\text{B} T}{E_\text{B,0}} \ln \left(\dfrac{\tau}{\tau_0}\right) \right) \quad .
\label{NeelArrhnenius}
\end{equation}
For the attempt time $\tau_0$ we assume $\tau_0 = 1$~ns. We obtain the fits presented as dashed lines in Fig. \ref{j_sw} (b) and (c). For $B_\text{x}=10$~mT we find a nucleation energy barrier of $E_\text{B,0,WHf} \approx 0.65$~eV for the W$_{66}$Hf$_{34}$ sample and $E_\text{B,0,Ta} \approx 0.78$~eV for the Ta reference. The deviation for short pulses from Eq. \ref{NeelArrhnenius} can be explained by slow domain wall propagation: While the current density is large enough to nucleate an opposing domain, it is not large enough to drive the domain through the line bar, such that higher current density is needed to achieve full switching in the time given by the current pulses.\\

For better understanding of the origin of the low switching current density we measure the velocity of domain walls $v_\text{DW}$ as a function of the driving force, which is the the applied current density in our experiement. It can be generally divided in two regimes, the creep and the flow regime, separated by a region in which the depinning of the domain walls takes place\cite{Jeudy2016,DiazPardo2017,Jeudy2018}. In the flow regime, the velocity shows a linear dependence on the driving force while in the creep regime, the motion of domain walls is thermally activated and the velocity can be described by the creep law
\begin{equation}
v_\text{DW} = v_\text{dep} \cdot \exp \left( - \dfrac{T_\text{dep}}{T} \cdot \left[ \left(\dfrac{j}{j_\text{dep}}\right)^{-1/4} -1\right] \right)
\label{v_DW} \quad .
\end{equation}
\begin{figure}[b]
	\includegraphics[width=8.6cm]{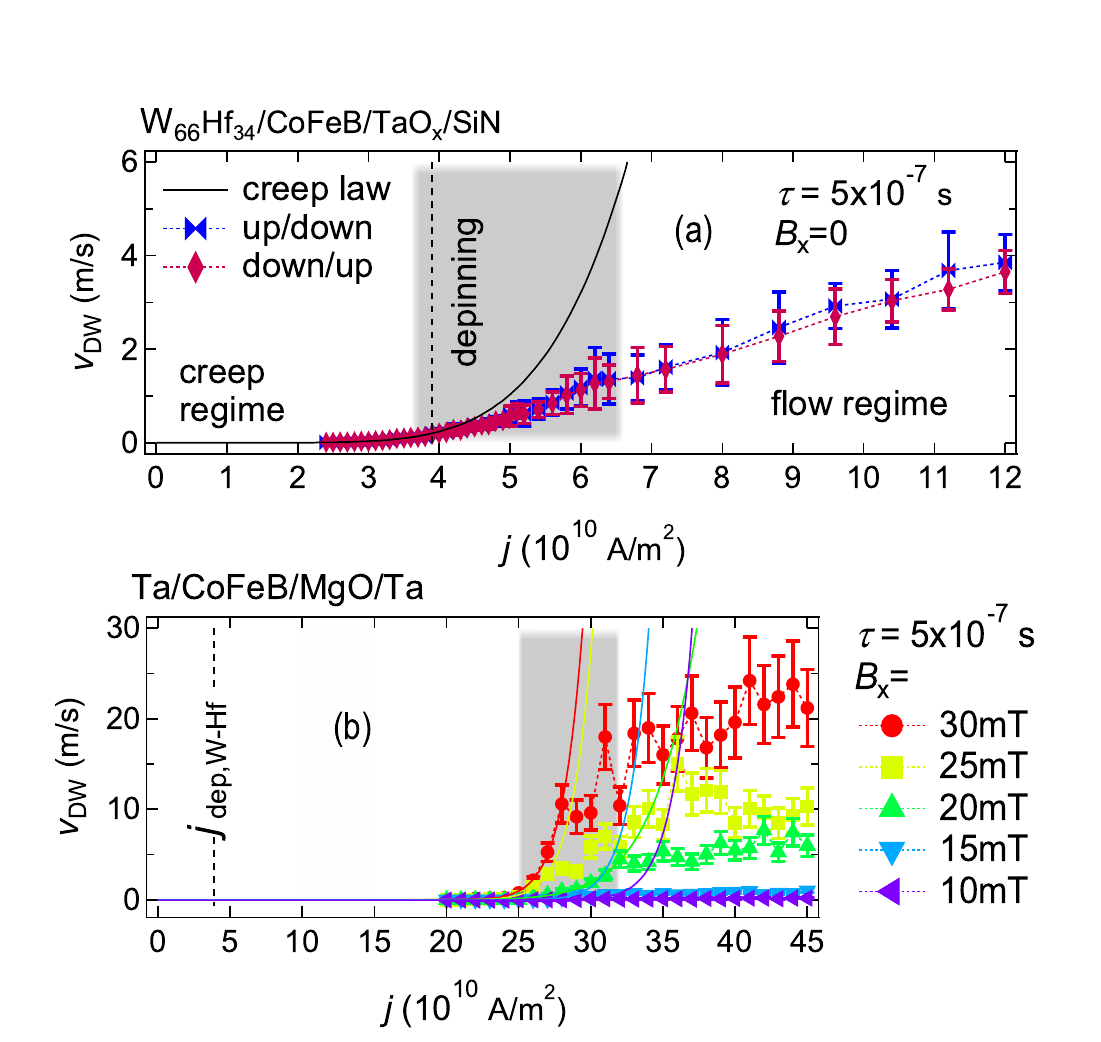}
	\caption{\label{j_dep} Domain wall velocity as a function of the current density in (a) W$_{66}$Hf$_{34}$/CoFeB/TaO$_\text{x}$/SiN and (b) Ta/CoFeB/MgO/Ta for different in-plane fields. The highlighted area marks the region of depinning.}
\end{figure}
\begin{figure}[b]
	\includegraphics[width=7cm]{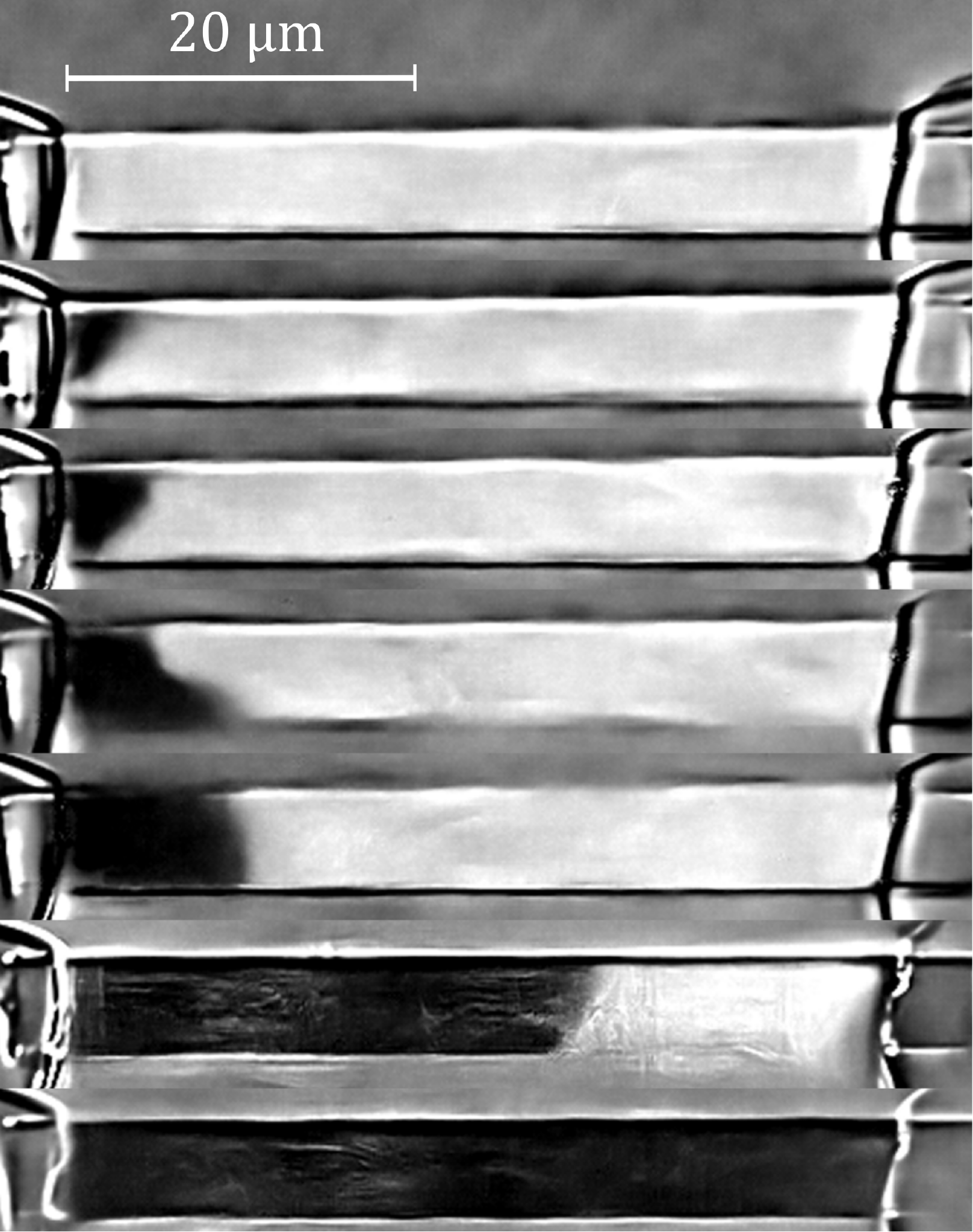}
	\caption{\label{switching_pictures} Differential Kerr images depicting CIMS in W$_{66}$Hf$_{34}$/CoFeB/TaO$_\text{x}$/SiN. Single current pulses with $j=2\times 10^{10}$\,A/m$^2$ and $\tau = 1\times 10^{-4}$\,s were applied with an in-plane field of $B_\text{x}=20$\,mT. }
\end{figure}

Here, $j_\text{dep}$ denotes the depinning current density and $T_\text{dep}/T = E_\text{dep}/k_\text{B} T = \Delta_\text{dep}$ corresponds to the thermal stability factor of the system \cite{Khvalkovskiy2013}. For the W$_{66}$Hf$_{34}$ sample the measurements were performed in the $50 \times 6$\,$\mu$m$^2$ line bars by applying current pulses with $\tau=5\times 10^{-7}$\,s and varying number $N$. CIDWM is observed for zero assist field, indicating that the domain walls are of a partial Néel-type due to the presence of Dzyaloshinskii-Moriya interaction (DMI)\cite{Franken2014}. Without an external magnetic field, up/down and down/up domain walls move in the same direction, indicating that they have the same chirality. The motion of up/down and down/up domain walls was analysed separately. The resulting domain wall velocity for zero assist field, presented in Fig. \ref{j_dep}(a), shows the expected change in slope in the grey shaded area and, therefore, allows for a distinction between the creep and flow regime. By fitting the creep law (Eq. \ref{v_DW}) to the data obtained for low current densities up to the grey shaded area we find a low depinning current density of $j_\text{dep} = (3.9 \pm 0.3) \times 10^{10}$\,A/m$^2$, marked with the dashed black line, and a thermal stability factor of $\Delta_\text{dep}=27\pm4$. We define the depinning current density as the value $j$ where the creep law deviates from the measured data by at least one standard error. Interpolating the CIMS results in Fig. \ref{j_sw}\,(b) to $\tau = 5 \times 10^{-7}$\,s for $B_\text{x}=10$\,mT we obtain the corresponding switching current density $j_\text{sw} \approx 2.5\times 10^{10}$\,A/m$^2$, i.e. nearly identical to the depinning current density for the same pulse width. We note that $B_\text{x}=10$\,mT is approximately the threshold value for observation of CIMS, which indicates that only for larger $B_x$ the homochirality of up/down and down/up domain walls is broken. This allows to quantify the strength of the DMI from an individual analysis of $v_\text{DW}$ for up/down and down/up domain walls as a function of $B_x$\cite{Soucaille2016,Vanatka2015}. A minimum of $v_\text{DW}$ is expected for a particular $B_\text{x}$, where the sign of this field is opposite for the up/down and down/up domain walls (not shown). Here, the effective longitudinal DMI field is exactly compensated, resulting in Bloch-type domain walls at one side of a domain and, therefore, vanishing $B_\mathrm{DL}$. From the threshold field $B_\text{x} = B_\text{DMI}$, the effective DMI constant
\begin{equation}
D = B_\text{DMI} M_\text{s} \Delta_\mathrm{DW}
\end{equation}
can be calculated \cite{Soucaille2016,Vanatka2015}. Here, $\Delta_\mathrm{DW} = \sqrt{A/K_\text{eff}}$ is the domain wall width parameter in which $A \approx 20$\,pJ/m denotes the exchange constant \cite{Chaurasiya2016} and $K_\text{eff}= B_\text{ani} M_s/2$ is the effective uniaxial anisotropy \cite{Vanatka2015}. With a threshold value $B_\text{x}\approx 10$\,mT $=B_\text{DMI}$ and $\Delta_\mathrm{DW} \approx 22$\,nm, a DMI constant of $D \approx 0.15$~mJ/m$^2$ is obtained. 

For $j_\text{dep} = 3.9 \times 10^{10}$~A/m$^2$ and $\tau = 5\times 10^{-7}$~s, we find a depinning energy barrier of $E_\text{dep} \approx 0.69$~eV which is comparable to the corresponding energy barrier $E_\text{B,0} \approx 0.65$~eV for nucleation of an opposing domain. This results suggests that, as soon as an opposing domain is nucleated, it can expand almost undisturbed in the line bar at the same current density. Observation of the CIMS process in the W$_{66}$Hf$_{34}$ line bars in the differential Kerr image support this interpretation: As can be seen in Fig. \ref{switching_pictures}, we find that the magnetization reversal happens via nucleation of a single domain which then quickly expands upon application of current pulses. Thus, the switching is dominated by the motion of domain walls rather than by the nucleation process \cite{Prommier1990} and, therefore, just the domain wall velocity is the limiting factor for CIMS.

In the Ta reference sample, CIDWM is only observed under application of longitudinal fields, indicating Bloch-type domain walls \cite{Franken2014}, where the magnetic moments in the domain walls are aligned parallel to spin direction $\bm{\sigma}$ of the spin current. In order for a spin current to cause a DL field $\bm{B}_\text{DL} = \bm{\sigma} \times \bm{m}$, the magnetic moments have to be tilted by an external field to have a magnetization component perpendicular to $\bm{\sigma}$. For this experiment, thinner and longer line bars had to be used to prevent nucleation of domain walls at the line edges due to the Oersted field and, on the other hand, to be able to observe the faster domain wall propagation. The resulting domain wall velocity, measured in line bars with a geometry of $100\times 2$\,$\mu$m$^2$ for different longitudinal fields and with $\tau=5\times 10^{-7}$\,s, is presented in Fig. \ref{j_dep}(b). Again, we observe a change in slope in the grey shaded area.

Larger in-plane assist fields force the magnetization within the domain walls to be parallel to the applied current, resulting in a larger $B_\mathrm{DL}$ and higher DW velocities. The depinning current density is obtained from fitting the creep law to the 10\,mT measurement, as this field equals the DMI effective field of the W$_{66}$Hf$_{34}$ sample. We find $j_\text{dep} = (2.9 \pm 0.3) \times 10^{11}$\,A/m$^2$ and a thermal stability factor $\Delta_\text{dep} = 105\pm 7$. The switching current density for comparable pulse parameters is obtained by extrapolation of the data in Fig. \ref{j_sw}\,(c). For $B_x = 10$\,mT we obtain $j_\text{sw} \approx 2.5 \times 10^{11}$\,A/m$^2$, which is almost the same value as the depinning current density. Using Eq. \ref{heatmapequation} we calculate the corresponding temperature rise of $\Delta T \approx 29$~K for the $100 \times 2$~$\mu$m$^2$ line bar and a resulting depinning energy barrier of $E_\text{dep} \approx 2.9$~eV, which is considerably larger than the according nucleation barrier $E_\text{B,0} \approx 0.78$~eV. This suggests that, in contrast to the W$_{66}$Hf$_{34}$ sample, in the Ta reference sample the switching is dominated by the nucleation of various domains and the pinning of the domain walls is the limiting factor for the switching. This conclusion is supported by observations of the nucleation process of opposing domains in both systems (Fig.\,\ref{switching_planefilm} and \ref{planefilm_Ta}).

A comparison of the results of both samples also emphasizes a direct correlation between $j_\text{dep}$ and $j_\text{sw}$, as both current densities are of the same order of magnitude, respectively. Additionally to the lower depinning and switching current densities in the W$_{66}$Hf$_{34}$-based sample, we observe a significantly lower thermal stability factor $\Delta_\text{dep}$, which is in good agreement with earlier investigations of the correlation between the thermal stability and switching current density in CIMS and CIDWM experiments with submicron stripes, where pinning results from edge effects \cite{Kab-Jin2014}. 

Measurements of the DL and FL effective fields were performed in both samples to ensure that the significantly lower switching and depinning current densities in W$_{66}$Hf$_{34}$ are not the result of a much larger SOT efficiency. For the Ta reference, we find identical SOT efficiencies for both in-plane and out-of-plane measurement schemes. Furthermore, DL and FL values are very similar, incidentally, where the weighted mean is $\xi_\text{DL} \approx \xi_\text{FL} \approx -0.056 \pm 0.004$. In the W$_{66}$Hf$_{34}$ sample, the out-of-plane measurements scheme yields $\xi_\text{DL} \approx -0.138 \pm 0.063$ and $\xi_\text{FL} \approx -0.017 \pm 0.063$. From the in-plane configuration, we obtain $\xi_\text{DL} \approx -0.143 \pm 0.015$ and $\xi_\text{FL} \approx 0.092 \pm 0.034$. Again, the DL torques are very similar in both schemes, whereas the FL efficiencies are clearly different. The efficiency of the DL torque, which gives rise to the current-induced motion of the domain walls is approximately 2.5 times larger in the W$_{66}$Hf$_{34}$-based sample. However, this is not sufficient to explain the factor of up to 40 between the switching and depinning current densities of the two sample types. The main difference between the samples is thus clearly the pinning strength, which is very weak in the W$_{66}$Hf$_{34}$-based sample.

\section{Conclusion}

\begin{table}[b]
	\centering
	\begin{tabular}{r|c|c|}
		& W$_{66}$Hf$_{34}$/CoFeB/TaO$_\text{x}$ & Ta/CoFeB/MgO/Ta \\\hline
		$D_\text{z}$ & - & $\approx 6$\,nm\\
		$\overline{d}$ & $\approx 145\,\mu$m & $\approx 10\,\mu$m\\
		$j_\text{sw}$ & $\approx 10^9 - 10^{10}$\,A/m$^2$ & $\approx 10^{11}$\,A/m$^2$\\
		$j_\text{dep}$ & $\approx 10^{10}$\,A/m$^2$ & $\approx 10^{11}$\,A/m$^2$ \\
		$\Delta_\text{dep}$& $\approx 27$ & $\approx 105$\\
	\end{tabular}
	\caption{\label{summarytable} Comparison of the evaluated key points of both investigated systems: length of coherent scattering $D_\text{z}$, pinning site mean distance $\overline{d}$, switching and depinning current densities $j_\text{sw}$ and $j_\text{dep}$ and the thermal stability factor $\Delta_\text{dep}$. For better conciseness, the values here are an estimation of the order of magnitude or rounded.}
\end{table}

In summary, we investigated the current-induced magnetization switching in an all-amorphous W$_{66}$Hf$_{34}$/CoFeB/TaO$_\text{x}$ sample stack, obtaining ultra-low switching currents densities as low as $3\times 10^{9}$\,A/m$^2$. In order to understand the origin of this ultra-low switching current density we evaluated various characteristics of the sample, some of which are summarised in Tab. \ref{summarytable}. Observation of domain wall motion reveals a depinning current density in the same order of magnitude with a ratio of  $j_\text{dep}/j_\text{sw} \approx 1.6$. A comparison with a Ta/CoFeB/MgO/Ta reference sample yields $j_\text{dep}/j_\text{sw} \approx 1.2$, emphasising that this ratio is universally close to unity \cite{Kim2014}. This result indicates a correlation between the switching and depinning current density, where the switching is limited by the propagation of domain walls, rather than by their nucleation. We support this conclusion with an evaluation of the energy barriers for the nucleation and depinning $E_\text{B,0}$ and $E_\text{dep}$, and observations of the nucleation and expansion of opposing domains in both systems. The latter reveals strongly differing domain structures with a mean distance of pinning sites $\overline{d}$ of at least one order of magnitude larger in the W$_{66}$Hf$_{34}$-based sample. Based on X-ray diffraction pattern we ascribe this reduced pinning to the lack of a crystal structure and in particular to the lack of grain boundaries and associated inhomogeneities of the magnetic parameters. The ultra-low switching current densities we obtained for W$_{66}$Hf$_{34}$ therefore suggest that the all-amorphous character of the sample results in reduced pinning. This combination of a low pinning and the potentially large spin Hall angle in all-amorphous heavy metals, given by the intrinsic spin Hall effect due to the high resistivity, could be interesting for further investigations in the context of magnetization switching.

\begin{acknowledgments}
The authors thank G. Reiss for making available the laboratory equipment. This work was partially supported by the Deutsche Forschungsgemeinschaft under sign ME 4389/2-1.
\end{acknowledgments}

\section*{References}




\end{document}